\documentstyle[sprocl,epsf]{article}
\def\lsim{\mathrel{\rlap{\lower4pt\hbox{\hskip1pt$\sim$}}
    \raise1pt\hbox{$<$}}}         
\def\gsim{\mathrel{\rlap{\lower4pt\hbox{\hskip1pt$\sim$}}
    \raise1pt\hbox{$>$}}}         

\input psfig.sty

\def\Journal#1#2#3#4{{#1} {\bf #2}, #3 (#4)}

\def\AJP{\em Am. J. Phys.}

\def\NPB{{\em Nucl. Phys.} B}
\def\PLB{{\em Phys. Lett.}  B}
\def\PRL{\em Phys. Rev. Lett.}
\def\PRD{{\em Phys. Rev.} D}
\def\PRC{{\em Phys. Rev.} C}
\def\NJP{\em New J. Phys.}

\def\ZP{\em Z. Phys.}

\def\AJ{\em Ap. J.}

\def\RMP{\em Rev. Mod. Phys.}
\def\S{\em Science}

\def\SJNP{\em Sov. J. Nucl. Phys.}
\def\PPNP{\em Prog. Part. Nucl. Phys.}

\def\be{\begin{equation}}
\def\ee{\end{equation}}
\def\bea{\begin{eqnarray}}
\def\eea{\end{eqnarray}}

\begin{document}

\title{NEUTRINO OSCILLATIONS AND THE SOLAR NEUTRINO PROBLEM}

\author{W. C. HAXTON}

\address{Institute for Nuclear Theory, Box 351550, and
Department of Physics, Box 351560\\
University of Washington, Seattle, WA 98195, USA\\
E-mail: haxton@phys.washington.edu}


\maketitle\abstracts{ I describe the current status of the solar
neutrino problem, summarizing the arguments that its resolution
will require new particle physics.  The phenomenon of matter-enhanced
neutrino oscillations is reviewed.  I consider the implications
of current experiments -- including the SuperKamiokande 
atmospheric neutrino and LSND measurements -- and the need for
additional constraints from SNO and other new detectors.}

\section{Introduction}
Part of the interest in neutrino astrophysics has to do with
the fascinating interplay between nuclear and particle physics issues ---
e.g., whether neutrinos are massive and undergo flavor oscillations,
whether they have detectable electromagnetic moments, etc. ---
and astrophysical phenomena, such as the clustering of matter
on large scales, the processes responsible for the synthesis
of nuclei, the mechanism for core-collapse supernovae,
and the evolution of our sun.  This summary addresses one of the
oldest problems in neutrino astrophysics, the 30-year puzzle of
the missing solar neutrinos.  This puzzle grew out of attempts 
to test the standard theory of main sequence stellar evolution,
but has now led to speculations about physics beyond the standard
model of electroweak interactions.  I will describe the work
that defined the solar neutrino problem, the likelihood that 
its resolution is connected with massive neutrinos, and the 
hopes we have for future experiments.

\section{Open Questions in Neutrino Physics~\protect\cite{holstein}}
The existence of the neutrino was first suggested by Wolfgang
Pauli in a private letter dated December, 1930.  The motivation
was to solve an apparent problem with energy conservation in
nuclear $\beta$ decay: the observable particles in the final
state (the daughter nucleus and emitted electron) carried less
energy than that released in the nuclear decay.  Pauli suggested
that an unobserved particle, the neutrino, accompanied the decay
and accounted for the missing energy.

A number of important developments followed Pauli's suggestion.
In 1934 Fermi \cite{fermi} suggested a theory of $\beta$ decay that was 
modeled after electromagnetism, except that there was no analog
of the electromagnetic field: the interaction occurred at a 
point.  (Apart from the missing aspect of parity violation, 
this was the correct reduction of today's standard model to an
effective theory.)  In 1934 Bethe and Critchfield described
the role of $\beta$ decay in thermonuclear reaction chains
powering the stars
\[ (A,Z) \rightarrow (A,Z-1) + e^+ + \nu_e \]
thus predicting that our sun produces an enormous neutrino
flux.  In 1956 Cowan and Reines \cite{cowan} succeeded in measuring neutrinos
emitted by a reactor through the reaction
\[ \bar{\nu}_e + p \rightarrow n + e^+, \]
exploiting the positron and neutron coincidence. (The neutron was 
detected by $(n,\gamma)$ on a Cd neutron poison.)  In 1957
the weak force mediating neutrino interactions was found to
violate parity maximally.  Later experiments found that the
$\nu_e$ was replicated twice more in nature -- the $\nu_\mu$ and
$\nu_\tau$ -- each accompanying a distinct charged lepton,
\[ \nu_e \leftrightarrow e^-~~~\nu_\mu \leftrightarrow \mu^-~~~\nu_\tau \leftrightarrow \tau^-. \]
Finally all of this physics was
embodied in the standard electroweak model, out of which came
the prediction of a new neutral interaction mediating neutrino
scattering.

Despite all of this progress, a remarkable number of questions 
remain.  We now believe neutrinos are massive, but still have no
measurement of an absolute neutrino mass (only mass differences).
Many models attribute the puzzle of neutrino mass --- why these
neutrinos are so much lighter than other standard model particles ---
to scales well beyond the standard model, but we lack independent
experimental tools for probing these scales.  We do not now the
particle-antiparticle conjugation properties of neutrinos:
because they carry no standard model charges, both the Dirac
(distinct antiparticle) and Majorana (no distinction between
particle and antiparticle) possibilities are open.  An associated
question is the existence of nonzero electromagnetic moments:
magnetic, charge radius, anapole, and electric dipole.  No nonzero
moment has been measured.

Finally, there are many questions about
the role of neutrinos in astrophysics and cosmology.  We suspect
cosmic background neutrinos contribute to dark matter and may
influence large-scale structure formation.  However direct 
experimental attempts to measure background neutrinos have failed by
many orders of magnitude to reach the expected density.
Type II supernovae convert approximately 99\% of the energy 
released in the infall into neutrinos of all flavors.  Yet only 
$\bar{\nu}_e$s were detected from SN1987A.  Supernova modelers predict
that neutrinos play an essential role in the explosion mechanism and
in the associated nucleosynthesis, yet there is disagreement 
about the success of neutrino-driven 
explosions.  Finally, there is great interest in mounting searches
for very high energy astrophysical neutrinos that might be
associated with active galactic nuclei, gamma ray bursts, etc.

Given all of these open questions, 70 years after Pauli's 
original suggestion, it would be nice to have a few more answers.
There is every indication that some answers will come with the
resolution of the solar neutrino puzzle.

\section{The Standard Solar Model~\protect\cite{bbp98,tcl}}
Solar models trace the evolution of the sun over the past
4.7 billion years of main sequence burning, thereby predicting
the present-day temperature and composition profiles of the solar
core that govern neutrino production.  Standard solar models (SSMs) 
share four basic assumptions:
    
\noindent
* The sun evolves in hydrostatic equilibrium, maintaining
a local balance between the gravitational force and the pressure
gradient.  To describe this condition in detail, one must 
specify the equation of state as a function of temperature,
density, and composition.

\noindent
* Energy is transported by radiation and convection.  While
the solar envelope is convective, radiative transport dominates
in the core region where thermonuclear reactions take place.
The opacity depends sensitively on the solar composition, particularly
the abundances of heavier elements.

\noindent
* Thermonuclear reaction chains generate solar energy.
The standard model predicts that over 98\% of this energy
is produced from the pp chain conversion of four protons into $^4$He
(see Fig. 1)
\begin{equation}
          4p \rightarrow ^4\mathrm{He} + 2e^+  + 2 \nu_e 
\end{equation}
with proton burning through the CNO cycle contributing the remaining 2\%.  The sun is
a large but slow reactor: the core temperature, $T_c \sim  1.5 \cdot 10^7$ K,
results in typical center-of-mass energies for reacting particles
of $\sim$ 10 keV, much less than the Coulomb barriers inhibiting
charged particle nuclear reactions.  Thus reaction cross
sections are small: in most cases laboratory measurements are
only possible at higher energies, so that cross section data must be
extrapolated to the solar energies of
interest.
    
\noindent
* The model is constrained to produce today's solar
radius, mass, and luminosity.  An important assumption of
the standard model is that the sun was highly convective,
and therefore uniform in composition, when it first
entered the main sequence.  It is furthermore assumed
that the surface abundances of metals (nuclei with A $>$ 5)
were undisturbed by the subsequent evolution, and thus
provide a record of the initial solar metallicity.  The
remaining parameter is the initial $^4$He/H ratio, which
is adjusted until the model reproduces the present solar
luminosity in today's sun.  The resulting
$^4$He/H mass fraction ratio is typically 0.27 $\pm$ 0.01,
which can be compared to the big-bang value of 0.23 $\pm$ 0.01.  
Note that the sun was formed from previously processed
material.

\begin{figure}[htb]
\psfig{bbllx=0.5cm,bblly=4.0cm,bburx=18cm,bbury=18.5cm,figure=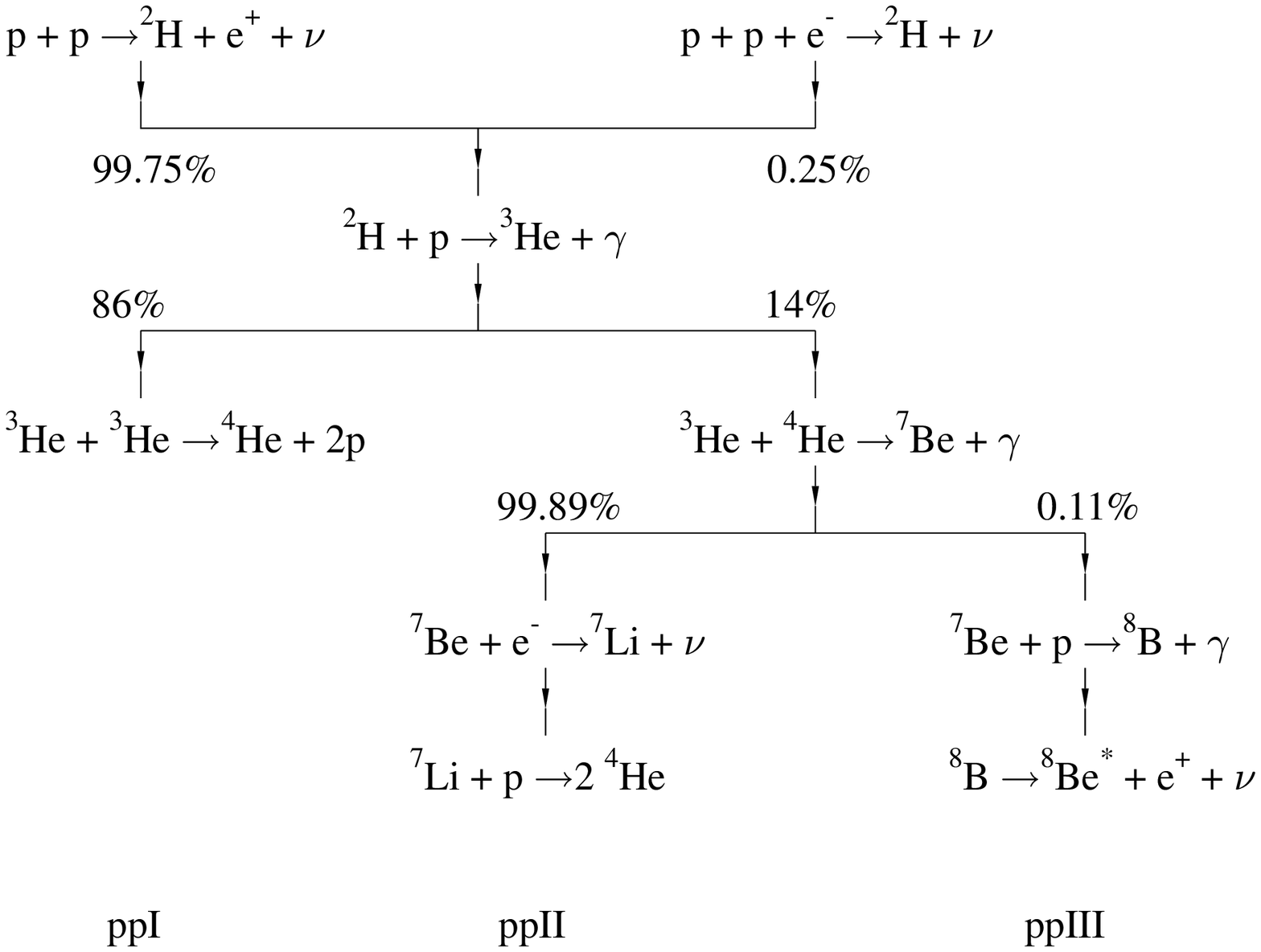,height=3.3in}
\caption{The solar pp chain.}
\end{figure}
  
The model that emerges is an evolving sun.  As the
core's chemical composition changes, the opacity and
core temperature rise, producing a 44\% luminosity increase
since the onset of the main sequence.
The temperature rise governs the competition between the
three cycles of the pp chain: the ppI cycle dominates below
about 1.6 $\cdot 10^7$ K; the ppII cycle between
(1.7-2.3) $\cdot 10^7$K; and the ppIII above 2.4 $\cdot 10^7$K.
The central core temperature of today's SSM is about 
1.55 $\cdot 10^7$K.

The competition between the cycles determines the pattern
of neutrino fluxes.  Thus one consequence of the thermal
evolution of our sun is that the $^8$B neutrino
flux, the most temperature-dependent component, proves to
be of relatively recent origin: the predicted flux 
increases exponentially with a doubling period of about
0.9 billion years.

A final aspect of SSM evolution is the formation of composition
gradients on nuclear burning timescales.  Clearly there is a
gradual enrichment of the solar core in $^4$He, the ashes 
of the pp chain.  Another element, $^3$He, can be considered a
catalyst for the pp chain, being produced and then consumed,
and thus eventually reaching some equilibrium abundance.
The timescale for equilibrium to be established as well
as the final equilibrium abundance are both sharply
decreasing functions of temperature, and therefore increasing
functions of the distance from the center of the core.
Thus a steep $^3$He density gradient is
established over time.

The SSM has had some notable successes.  From helioseismology \cite{fr}
the sound speed profile $c(r)$ has been very accurately 
determined for the outer 90\% of the sun, and is in excellent
agreement with the SSM.  Such studies verify important predictions
of the SSM, such as the depth of the convective zone.
However the SSM is not a complete model in that it does 
not explain all features of solar structure, such as the 
depletion of surface Li by two orders of magnitude.  This is 
usually attributed to convective processes that operated at
some epoch in our sun's history,
dredging Li to a depth where burning takes place.
  
The principal neutrino-producing reactions of the pp chain
and CNO cycle are summarized in Table 1.  The first six reactions
produce $\beta$ decay neutrino spectra having allowed shapes with 
endpoints given by E$_\nu^{\rm max}$.  Deviations from an allowed spectrum
occur for $^8$B neutrinos because the $^8$Be final state is a broad 
resonance.  The last two reactions
produce line sources of electron capture neutrinos, with
widths $\sim$ 2 keV characteristic of the temperature of the solar core.
Measurements of the  pp, $^7$Be, and $^8$B neutrino fluxes will
determine the relative contributions of the  ppI,  ppII, and
ppIII cycles to solar energy generation.  As discussed above, 
and as later illustrations will show more clearly, 
this competition is governed in large
classes of solar models by a single parameter, the central
temperature $T_c$.  The flux predictions of 
the 1998 calculations of Bahcall, Basu, and Pinsonneault~\cite{bbp98} (BP98) 
and of Brun, Turck-Chieze and Morel~\cite{tcl} are included
in Table 1.

\begin{table}[t]
\caption{Solar neutrino sources and the flux predictions of the
BP98 and Brun/Turck-Chieze/Morel SSMs in cm$^{-2}$s$^{-1}$.}
\vspace{0.2cm}
\begin{center}
\begin{tabular}{|c|c|c|c|}
\hline
 & & & \\
Source & E$_\nu^{max}$ (MeV) & BP98 & BTCM98 \\
& & & \\
\hline
& & & \\
p + p $\rightarrow ^2$H + e$^+ + \nu$ & 0.42 & 5.94E10 & 5.98E10 \\
$^{13}$N $\rightarrow ^{13}$C + e$^+ + \nu$ & 1.20 & 6.05E8 & 4.66E8 \\
$^{15}$O $\rightarrow ^{15}$N + e$^+ + \nu$ & 1.73 & 5.32E8 & 3.97E8 \\
$^{17}$F $\rightarrow ^{17}$O + e$^+ + \nu$ & 1.74 & 6.33E6 & \\
$^8$B $\rightarrow ^8$Be + e$^+ + \nu$ & $\sim$ 15 & 5.15E6 & 4.82E6 \\
$^3$He + p $\rightarrow ^4$He + e$^+ + \nu$ & 18.77 & 2.10E3 & \\
$^7$Be + e$^- \rightarrow ^7$Li + $\nu$ & 0.86 (90\%) & 4.80E9 & 4.70E9 \\
 & 0.38 (10\%) & & \\
p + e$^-$ + p $\rightarrow ^2$H + $\nu$ & 1.44 & 1.39E8 & 1.41E8 \\
 & & & \\
\hline
\end{tabular}
\end{center}
\end{table}
  
\section{Solar Neutrino Experiments and their Implications}
  
The first solar neutrino results were announced by Ray Davis Jr. 
and his Brookhaven collaborators in 1968, more than 30 years ago \cite{davis}.
Located deep within the Homestake Gold Mine in Lead, South Dakota,
the detector consists of a 100,000 gallon tank of C$_2$Cl$_4$.
Solar neutrinos are captured by
\[ ^{37}\mathrm{Cl}(\nu,e^-)^{37}\mathrm{Ar}. \]
As the threshold for this reaction is 0.814 MeV, the important
neutrino sources are the $^7$Be and $^8$B reactions.
The $^7$Be neutrinos excite just the Gamow-Teller (GT) transition to
the ground state, the strength of which is known from the 
electron capture lifetime of $^{37}$Ar.  The $^8$B neutrinos
can excite all bound states in $^{37}$Ar, including the
dominant transition to the isobaric analog state residing at an excitation
energy of 4.99 MeV.  The strength of excite-state GT transitions
can be determined from the $\beta$ decay $^{37}$Ca$(\beta^+)^{37}$K,
which is the isospin mirror reaction to $^{37}$Cl$(\nu,e^-)^{37}$Ar.
The net result is that, for SSM fluxes, 78\% of the capture
rate should be due to $^8$B neutrinos, and 15\% to $^7$Be
neutrinos.  The measured capture rate~\cite{lande} 2.56 $\pm 0.16 \pm 0.16$ SNU
(1 SNU = 10$^{-36}$ capture/atom/sec) is about one-third the 
SSM value.

Similar radiochemical experiments were begun in January, 1990,
and May, 1991, respectively, 
by the SAGE and GALLEX collaborations using a different 
target, $^{71}$Ga.  The special properties of this target include
its low threshold and an unusually strong transition to the ground state of 
$^{71}$Ge, leading to a large pp neutrino cross section (see Fig. 2).
The experimental capture rates are
$66.6 {}^{+6.8}_{-7.1} {}^{+3.8}_{-4.0}$ (SAGE) \cite{sage} and $77.5 \pm 6.2 {}^{+4.3}_{-4.7}$ SNU (GALLEX) \cite{gallex}.
The SSM prediction is about 130 
SNU~\cite{bahcallb}.  Most important, since the pp flux is directly constrained
by the solar luminosity in all steady-state models, there is
a minimum theoretical value for the capture rate of 79 SNU,
given standard model weak interaction physics.  Note there
are substantial uncertainties in the $^{71}$Ga cross section
due to $^7$Be neutrino capture to two excited states of unknown 
strength.  These uncertainties were greatly reduced by
direct calibrations of both detectors using $^{51}$Cr
neutrino sources.

\begin{figure}[htb]
\psfig{bbllx=0.0cm,bblly=4.0cm,bburx=16cm,bbury=22.5cm,figure=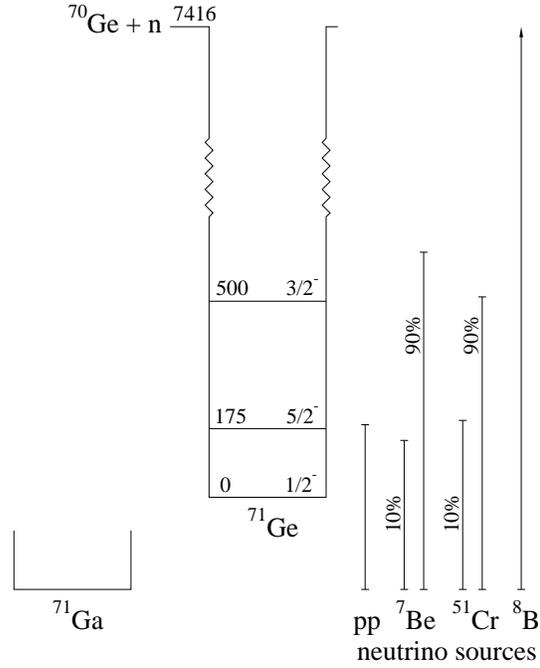,height=3.8in}
\caption{Level scheme for $^{71}$Ge showing the excited states
that contribute to absorption of pp, $^7$Be, $^{51}$Cr and
$^8$B neutrinos.}
\end{figure}
  
Experiments of a different kind, Kamiokande II/III and SuperKamiokande,
exploit water Cerenkov detectors to view solar
neutrinos event-by-event.  Solar neutrinos scatter off electrons,
with the recoiling electrons producing the Cerenkov radiation
that is then recorded in surrounding phototubes.  
Thresholds are determined by background rates; SuperKamiokande
is currently operating with a trigger at approximately six MeV.
The initial experiment, Kamiokande II/III, found a flux of
$^8$B neutrinos of (2.80 $\pm 0.19 \pm 0.33) \cdot 10^6$/cm$^2$s after
about a decade of measurement \cite{k}.  Its much larger successor
SuperKamiokande, with a 22.5 kiloton fiducial volume,
yielded the result $(2.42 \pm 0.04 \pm 0.06) \cdot 10^6$/cm$^2$s
after the first 825 days of measurements \cite{sk}.  This is about
48\% of the SSM flux.  This result continues to improve in accuracy. \\
  
These results can be combined to limit the principal solar
neutrino fluxes, under the assumption that no new particle physics
distorts the spectral shape of the pp and $^8$B neutrinos.
One finds
\begin{eqnarray}
\phi (pp) & \sim & 0.9 \, \phi^{\rm {SSM}} (pp)\nonumber \\
\phi (^7{\rm {Be}}) & \sim & 0 \nonumber\\
\phi (^8 {\rm B}) & \sim & 0.47 \, \phi^{\rm {SSM}} (^8{\rm B}).  
\end{eqnarray}
A reduced $^8$B neutrino flux can be produced by lowering
the central temperature of the sun somewhat, as $\phi(^8$B)$\sim T_c^{18}$.  However, such
an adjustment, either by varying the parameters of the SSM or by
adopting some nonstandard physics, tends to push the $\phi (^7$Be)/$\phi(^8$B)
ratio to higher values rather than the low one of eq. (12),
\begin{equation}
{\phi (^7{\rm{Be}}) \over \phi(^8 {\rm B})} \sim T_c^{-10}.
\end{equation}
Thus the observations seem difficult to reconcile with plausible
solar model variations: one observable ($\phi(^8$B)) requires a cooler
core while a second, the ratio $\phi(^7$Be)/$\phi(^8$B), requires a hotter one.

This physics was nicely illustrated by Castellani et al.~\cite{cast}.
These authors generated a series of nonstandard models by changing
the S-factor for the p+p reaction, modifying the core metalicity,
introducing weakly interacting massive particles as a new mechanism
for energy transport, etc.  The resulting core temperature $T_c$ and
neutrino fluxes were then determined, and the latter were plotted
as a function of the former.
The pattern that emerges is striking (see Fig. 3):
parameter variations producing the same value of 
$T_c$ produce
remarkably similar fluxes.  Thus 
$T_c$ provides an excellent
one-parameter description of standard model perturbations.
Figure 3 also illustrates the difficulty of producing a 
low ratio of $\phi(^7$Be)/$\phi(^8$B) when 
$T_c$ is reduced.  This result is consistent with
our earlier argument and shows that even extreme
changes in quantities like the metallicity, opacities, or
solar age, cannot produce the pattern of fluxes deduced
from experiment (eq. (2)).   

\begin{figure}[htb]
\psfig{bbllx=0.3cm,bblly=4.0cm,bburx=14.5cm,bbury=24.0cm,figure=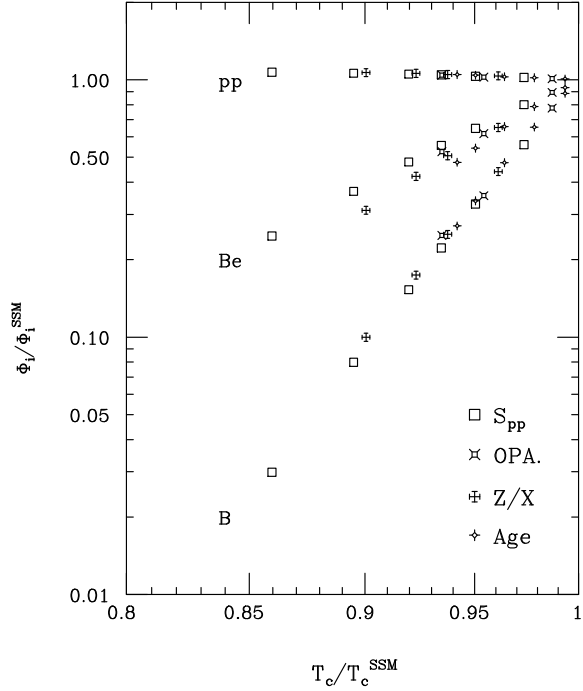,height=3.6in}
\caption{The responses of the pp, $^7$Be, and $^8$B neutrino fluxes
to the indicated variations in solar model input parameters,
displayed as a function of the resulting central temperature
$T_c$.  From Castellani et al.}
\end{figure}
  
Is it possible to change the solar model in a way that 
reduces the $^7$Be/$^8$B neutrino flux ratio?  It is appears
the answer is no in models where the nuclear reactions burn in
equilibrium with plausible cross sections.  However Cumming and
Haxton~\cite{cumming} pointed out that a possible exception was
a nonequilibrium model in which the solar core is mixed on the timescale of $^3$He evolution,
about $10^7$ years.  Thus the pp chain is prevented
from reaching equilibrium.  This suggestion has some physical
plausibility because it allows the sun to burn 
more efficiently, with a cooler core and enhanced ppI terminations.
The SSM $^3$He profile is known to be overstable, as was first
discussed by Dilke and Gough~\cite{gough}.  Also the possibility
of a persistent convective core powered by the $^3$He gradient
has been discussed in the literature.  (The SSM core is convective
for about $10^8$ years because of out-of-equilibrium burning of
the CNO cycle.)  A strong argument against a mixed core was
offered by Bahcall et al. \cite{bhelio}, who showed that homogenizing the 
core of the SSM led to very large changes in the helioseismology.
While this is a sobering result,
this test was not done in a self-consistent model.
However, there is work in progress
to test the helioseismology of a more realistic mixed core
model -- one where the $^4$He content of the core, the temperature
gradient, the nuclear reaction rates, and the luminosity are
handled consistently~\cite{guzik}.  If the helioseismology 
remains unacceptable, this will rule out the only solar model conjecture for
producing a reduced $^7$Be/$^8$B flux ratio, which is necessary
to produce fluxes closer to those observed.
   
However, there is a popular argument showing that no SSM change
can completely remove the discrepancy with experiment: if one assumes undistorted
neutrino spectra, no combination of pp, $^7$Be, and $^8$B
neutrino fluxes fits the experimental results well~\cite{karsten}.  In
fact, in an unconstrained fit, the required $^7$Be flux
is unphysical, negative by about 2.5$\sigma$.  This is clearly
a strong hint that one should 
look elsewhere for a solution!

The remaining possibility is new neutrino physics.
Suggested particle physics solutions of the solar
neutrino problem include neutrino oscillations, neutrino
decay, neutrino magnetic moments, and weakly interacting
massive particles.  Among these, the Mikheyev-Smirnov-Wolfenstein 
effect --- neutrino oscillations enhanced by matter
interactions --- is widely regarded as perhaps the most plausible.

\section{Neutrino Oscillations}
One odd feature of particle physics is that neutrinos,
which are not required by any symmetry to be massless, nevertheless
must be much lighter than any of the other known fermions.
For instance, the current limit on the $\overline{\nu}_e$ mass is $\lsim$ 5 eV.
The standard model requires neutrinos to be massless, but the
reasons are not fundamental.  Dirac mass terms $m_D$, analogous
to the mass terms for other fermions, cannot be constructed
because the model contains no right-handed neutrino fields.
Neutrinos can also have Majorana mass terms
\begin{equation}
\overline{\nu^c_L} m_L \nu_L ~~~ \mathrm{and} ~~~ \overline{\nu^c_R} m_R \nu_R 
\end{equation}
where the subscripts $L$ and $R$ denote left- and right-handed projections
of the neutrino field $\nu$, and the superscript $c$ denotes charge conjugation.
The first term above is constructed from left-handed fields, but can 
only arise as a nonrenormalizable effective interaction when
one is constrained to generate $m_L$ with the doublet scalar field of
the standard model.  The second term is absent from the standard  
model because there are no right-handed neutrino fields.

None of these standard model arguments 
carries over to the more general, unified theories that 
theorists believe will supplant the standard model.  
In the enlarged multiplets of extended models it is
natural to characterize the fermions of a single family,
e.g., $\nu_e$, e, u, d, by the same mass scale $m_D$.  Small neutrino
masses are then frequently explained as a result of the
Majorana neutrino masses.  In the seesaw mechanism,
\begin{equation}
M_\nu \sim \left(\begin{array}{cc}
0 & m_D \\
m^T_D & m_R \end{array}\right) .  
\end{equation}
Diagonalization of this matrix  
produces one light neutrino, $m_{\mathrm{light}}\sim m_D ({m_D \over m_R})$, and one 
unobservably
heavy, $m_{\mathrm{heavy}} \sim m_R$.  The factor ($m_D$/$m_R$) is the needed small
parameter that accounts for the distinct scale of neutrino
masses.  The masses for the $\nu_e, \nu_\mu$, and $\nu_\tau$ are then
related to the squares of the corresponding quark masses
$m_u$, $m_c$, and $m_t$.  Taking $m_R \sim 10^{16}$ GeV, a typical grand 
unification scale for models built on groups like SO(10), the seesaw mechanism gives the crude relation
\begin{equation}
m_{\nu_e}: m_{\nu_\mu}: m_{\nu_\tau} \leftrightarrow 2 \cdot 10^{-12}: 2 
\cdot 10^{-7}: 3 \cdot 10^{-3} \mathrm{eV}. 
\end{equation}
The fact that solar neutrino experiments can probe small
neutrino masses, and thus provide insight into possible new
mass scales $m_R$ that are far beyond the reach of direct
accelerator measurements, has been an important theme of
the field.
    
Consider for simplicity just two neutrino flavors.  The states
of definite mass are the states that diagonalize the free
Hamiltonian.  Similarly the weak interaction eigenstates are the
states of definite flavor, that is, the $\nu_e$ accompanies the
positron in $\beta$ decay, and the $\nu_\mu$ accompanies the
muon.  There is every reason to assume that these two bases
are not coincident, but instead are related by a nontrivial
rotation,
\begin{eqnarray}
|\nu_e\rangle  &=& \cos \theta_v |\nu_1\rangle  + \sin \theta_v|\nu_2 \rangle  \nonumber \\
|\nu_\mu\rangle &=& - \sin \theta_v |\nu_1 \rangle + \cos \theta_v |\nu_2 
\rangle 
\end{eqnarray}
where $\theta_v$ is the (vacuum) mixing angle. 
  
Consider a $\nu_e$ produced at time $t$=0 as a momentum eigenstate \cite{nauenberg}
\begin{equation}
|\nu(t=0)\rangle  = |\nu_e \rangle = \cos \theta_v |\nu_1\rangle  + \sin \theta_v|\nu_2 \rangle . 
\end{equation}
The resulting probability for measuring a $\nu_e$ downstream
then depends on $\delta m^2 = m_2^2-m_1^2$,
\begin{eqnarray}
P_{\nu_e} (t) &=& | \langle \nu_e | \nu(t) \rangle |^2 \nonumber \\ 
 &=& 1 - \sin^2 2 \theta_v \sin^2 \left({\delta m^2 t \over 4 
k}\right) \rightarrow 1 - {1 \over 2} \sin^2 2 \theta_v 
\end{eqnarray}
where the limit on the right is appropriate for large $t$.
(When one properly describes the neutrino state as a wave packet,
the large-distance behavior follows from the eventual separation
of the mass eigenstates.)  If the
the oscillation length
\begin{equation}
L_o = {4 \pi \hbar c E \over \delta m^2 c^4} 
\end{equation}
is comparable to or shorter than one astronomical unit, a 
reduction in the solar $\nu_e$ flux would be expected in terrestrial
neutrino oscillations.
  
The suggestion that the solar neutrino problem could
be explained by neutrino oscillations was first made by
Pontecorvo in 1958, who pointed out the analogy with $K_0 \leftrightarrow \bar 
K_0$     
oscillations.  From the point of view of particle physics, 
the sun is a marvelous neutrino source.  The neutrinos travel a long
distance and have low energies ($\sim$ 1 MeV), implying a sensitivity to
\begin{equation}
\delta m^2 \gsim 10^{-12} eV^2.
\end{equation}
In the seesaw mechanism, $\delta m^2 \sim m^2_2$, so neutrino masses as
low as $m_2 \sim 10^{-6}$ eV could be probed.  In contrast, terrestrial
oscillation experiments with accelerator or reactor
neutrinos are typically limited to $\delta m^2 \gsim 0.1 $ eV$^2$. 
(Planned long-baseline experiments, though, will soon push below
0.01 eV$^2$.)

From the expressions above one expects vacuum oscillations to affect
all neutrino species equally, if the oscillation length is small
compared to an astronomical unit.  This is somewhat in conflict
with the data, as we have argued that the $^7$Be neutrino flux
is quite suppressed.
Furthermore, there is a weak theoretical prejudice that $\theta_v$ should be
small, like the Cabibbo angle.
The first objection, however, can be circumvented in
the case of ``just so" oscillations where the oscillation 
length is comparable to one astronomical unit.
In this case the oscillation probability becomes sharply
energy dependent, and one can choose $\delta m^2$ to preferentially
suppress one component (e.g., the monochromatic $^7$Be neutrinos).
This scenario has been explored by several groups and
remains an interesting possibility.  However, the
requirement of large mixing angles remains.

\section{The Mikheyev-Smirnov-Wolfenstein Mechanism~\protect\cite{msw}}
  
In order to include matter effects, we first consider vacuum 
oscillations for the more general case
\begin{equation}
 |\nu(t=0)\rangle = a_e(t=0) |\nu_e \rangle + a_\mu(t=0) 
|\nu_\mu \rangle . 
\end{equation}
from which one easily calculates
\begin{equation}
i {d \over dx} \left( \matrix { a_{\textstyle e} \cr
a_{\textstyle \mu} \cr} \right) = {1 \over 4E} \left ( \matrix{
- \delta m^2 \cos 2 \theta_{\textstyle v}
~~~~~~~~~~~\delta m^2\sin
2\theta_{\textstyle v} \cr 
\delta m^2\sin 2 \theta_{\textstyle v} ~~~~~~~~~~~ 
\delta m^2
\cos 2\theta_{\textstyle v} \cr} \right) \left( \matrix {
a_{\textstyle e} \cr
a_{\textstyle \mu} \cr} \right) . 
\end{equation}
We have equated $x = t,$ that is, set $c$ = 1.
  
Mikheyev and Smirnov~\cite{msw} showed in 1985 that the
density dependence of the neutrino effective mass, a phenomenon
first discussed by Wolfenstein in 1978, could greatly enhance
oscillation probabilities: a $\nu_e$ is adiabatically transformed
into a $\nu_\mu$ as it traverses a critical density within the sun.
It became clear that the sun was not only an excellent 
neutrino source, but also a natural regenerator for cleverly
enhancing the effects of flavor mixing. 
   
While the original work of Mikheyev and Smirnov was
numerical, their phenomenon was soon understood analytically
as a level-crossing problem.  The vacuum oscillation evolution
equation changes in the presence of matter to
\begin{equation}
i {d \over dx} \left( \matrix { a_{\textstyle e} \cr
a_{\textstyle \mu} \cr} \right) = {1 \over 4E} \left ( \matrix{
2E \sqrt2 G_F \rho(x) - \delta m^2 \cos 2 \theta_{\textstyle v}
~~~~~~\delta m^2\sin
2\theta_{\textstyle v} \cr 
\delta m^2\sin 2 \theta_{\textstyle v} ~~~ -2E \sqrt2 G_F \rho(x) +
\delta m^2
\cos 2\theta_{\textstyle v} \cr} \right) \left( \matrix {
a_{\textstyle e} \cr
a_{\textstyle \mu} \cr} \right) 
\end{equation}
where G$_F$ is the weak coupling constant and $\rho (x)$ the solar
electron density.
The new contribution to the diagonal elements, $2 E \sqrt2 G_F \rho(x)$, 
represents the effective contribution to $M^2_\nu$  that arises 
from neutrino-electron scattering.  The indices of refraction
of electron and muon neutrinos differ because the former
scatter by charged and neutral currents, while the latter 
have only neutral current interactions.  The difference in
the forward scattering amplitudes determines the density-dependent
splitting of the diagonal elements of the new matter equation. 

It is helpful to rewrite this equation in a basis consisting of the light and heavy 
local mass eigenstates (i.e., the states that diagonalize the right-hand side 
of the equation),
\begin{eqnarray}
|\nu_L (x)\rangle &=& \cos \theta (x)|\nu_e\rangle - \sin \theta 
(x)|\nu_\mu\rangle \nonumber \\
|\nu_H(x)\rangle &=& \sin \theta (x)|\nu_e\rangle + \cos \theta (x)|\nu_\mu 
\rangle . 
\end{eqnarray}
The local mixing angle is defined by
\begin{eqnarray}
\sin 2 \theta (x)  &=& {\sin 2 \theta_{\textstyle v} \over \sqrt{X^2 (x) + \sin^2
2\theta_{\textstyle v}}} \nonumber \\
\cos 2\theta (x)  &=& {-X (x) \over \sqrt{X^2 (x) + \sin^2 2\theta_{\textstyle
v}}} 
\end{eqnarray}
where $X(x) = 2 \sqrt2G_F \rho(x) E/\delta m^2 - \cos 2\theta_{\textstyle v}$.
Thus
$\theta(x)$ ranges from $\theta_{\textstyle v}$ to $\pi/2$ as the density
$\rho(x)$ goes
from 0 to $\infty$. 

If we define
\begin{equation}
|\nu (x) \rangle = a_H(x)|\nu_H(x)\rangle + a_L(x)|\nu_L(x)\rangle,
\end{equation}
the neutrino propagation can be rewritten in terms of the local
mass eigenstates
\begin{equation}
i {d \over dx} \pmatrix{
a_H \cr
a_L \cr} = \pmatrix {
\lambda(x) & i \alpha (x) \cr
-i \alpha (x) & - \lambda (x) \cr }
\pmatrix
{a_H \cr
a_L }
\end{equation}
with the splitting of the local mass eigenstates determined by
\begin{equation}
2 \lambda (x) = {\delta m^2 \over 2E} \sqrt{X^2 (x) + \sin^2 2 
\theta_{\textstyle v}} 
\end{equation}
and with mixing of these eigenstates governed by the density gradient
\begin{equation}
\alpha (x) = \left({E \over \delta m^2}\right)
 \, {\sqrt2 \, G_F {d \over dx}
\rho(x)
\sin 2 \theta_{\textstyle v} \over X^2 (x) + \sin^2 2 \theta_{\textstyle v}}.
\end{equation}
The results above are quite interesting: the local mass eigenstates
diagonalize the matrix if the density is constant.  In such a limit,
the problem is no more complicated than our original vacuum
oscillation case, although our mixing angle is changed because of
the matter effects.  But if the density is not constant, the
mass eigenstates in fact evolve as the density changes.  This
is the crux of the MSW effect.
Note that the splitting achieves
its minimum value, ${\delta m^2 \over 2E} \sin 2 \theta_v$, at a critical density $\rho_c =
\rho (x_c)$
\begin{equation}
2 \sqrt2 E G_F \rho_c = \delta m^2 \cos 2 \theta_v 
\end{equation}
that defines the point where the diagonal elements of the original flavor matrix cross. 

Our local-mass-eigenstate form of the propagation equation can be trivially integrated if the splitting of the diagonal
elements is 
large compared to the off-diagonal elements,
\begin{equation}
\gamma (x) = \left|{\lambda (x) \over \alpha (x)}\right| = {\sin^2
2\theta_{\textstyle v} \over \cos
2\theta_{\textstyle v}} \, {\delta m^2 \over 2 E} \, {1 \over |{1 \over \rho_c}
{d \rho (x) \over
dx}|} {[X (x)^2 + \sin^2 2\theta_v]^{3/2} \over \sin^3 2\theta_v} \gg 1, 
\end{equation}
a condition that becomes particularly stringent near the crossing point,
\begin{equation}
\gamma_c = \gamma (x_c) = {\sin^2 2\theta_v \over \cos 2\theta_v} \, {\delta
m^2 \over 2 E} \, {1 \over \left|{1 \over \rho_c} {d \rho (x) \over dx}|_{x =
x_c}\right|} 
\gg 1. 
\end{equation}
The resulting adiabatic electron neutrino survival probability~\cite{bethe}, valid when
$\gamma_c \gg 1$, is
\begin{equation}
P^{\rm adiab}_{\nu_e} = {1 \over 2} + {1 \over 2} \cos 2 \theta_v \cos 2
\theta_i 
\end{equation}
where $\theta_i = \theta (x_i)$ is the local mixing angle at the density where
the neutrino was produced. 

\begin{figure}[htb]
\psfig{bbllx=1.2cm,bblly=2.0cm,bburx=18cm,bbury=14.5cm,figure=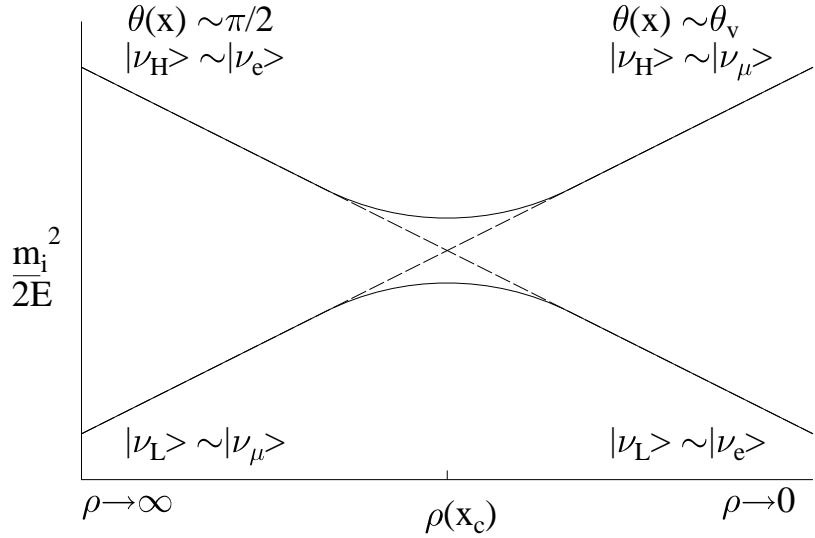,height=3.0in}
\caption{Schematic illustration of the MSW crossing.  The dashed 
lines correspond to the electron-electron and muon-muon diagonal
elements of the $M^2$ matrix in the flavor basis.  Their 
intersection defines the level-crossing density $\rho_c$.
The solid lines are the trajectories of the light and heavy
local mass eigenstates.  If the electron neutrino is produced 
at high density and propagates adiabatically, it will follow
the heavy-mass trajectory, emerging from the sun as a $\nu_\mu$.}
\end{figure}
  
The physical picture behind this derivation is illustrated
in Fig. 4.  One makes the usual assumption that, in vacuum,
the $\nu_e$ is almost identical to the light mass eigenstate,
$\nu_L(0)$, i.e., $m_1 < m_2$ and $\cos \theta_v \sim$ 1.  But as the density increases,
the matter effects make the $\nu_e$ heavier than the $\nu_\mu$, with $\nu_e 
\to \nu_H (x)$  as $\rho(x)$ becomes large.  That is, the mixing angle at high density 
rotates to $\pi/2$.  The special property of
the sun is that it produces $\nu_e$s at high density that then propagate to 
the vacuum where they
are measured.  The adiabatic approximation tells us that if
initially $\nu_e \sim \nu_H (x)$, the neutrino will remain on the heavy
mass trajectory provided the density changes slowly.
That is, if the solar density gradient is sufficiently gentle,
the neutrino will emerge from the sun as the heavy vacuum
eigenstate, $ \sim \nu_\mu$.  This guarantees nearly complete conversion
of $\nu_e$s into $\nu_\mu$s, producing a flux that cannot be detected
by the Homestake or SAGE/GALLEX detectors. 
   
But this does not explain the curious pattern of partial
flux suppressions coming from the various solar neutrino experiments.  The key to this is the behavior when
$\gamma_c \lsim$ 1.  Our expression for $\gamma(x)$ shows that the critical region
for nonadiabatic behavior occurs in a narrow region (for small $\theta_v$)
surrounding the crossing point, and that this behavior is 
controlled by the derivative of the density.  This suggests an
analytic strategy for handling nonadiabatic crossings: one
can replace the true solar density by a simpler (integrable!) two-parameter 
form that is constrained to reproduce the true density and its derivative at 
the crossing point $x_c$. Two convenient choices are the linear $(\rho (x) = a 
+ bx)$ and exponential $(\rho (x) = ae^{-bx})$ profiles.  As the density 
derivative at $x_c$ governs the nonadiabatic behavior, this procedure should 
provide an accurate description of the hopping probability between the local 
mass eigenstates when the neutrino traverses the crossing point.  The initial 
and ending points $x_i$ and $x_f$ for the artificial profile are then chosen 
so that $\rho(x_i)$ is the density where the neutrino was produced in the 
solar core and $\rho(x_f) = 0$ (the solar surface), as illustrated in in Fig. 5. 
Since the adiabatic result ($P_{\nu_e}^{\mathrm{adiab}}$) depends only on the local mixing angles 
at these points, this choice builds in that limit.  But our original flavor-basis equation can then be integrated 
exactly for linear and exponential profiles, with the results given in terms 
of parabolic cylinder and Whittaker functions, respectively.   

\begin{figure}[htb]
\psfig{bbllx=0.0cm,bblly=2.8cm,bburx=16cm,bbury=21.3cm,figure=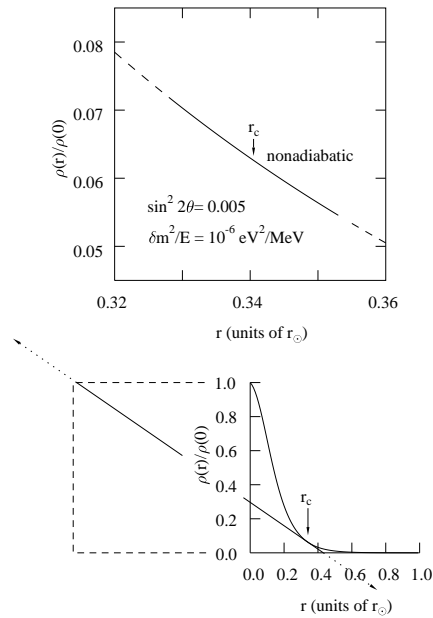,height=3.3in}
\caption{The top figure illustrates, for one choice of sin$^2 2\theta$
and $\delta m^2$, that the region of nonadiabatic propagation
(solid line) is usually confined to a narrow region around the
crossing point $r_c$.  In the lower figure, the solid lines
represent the solar density and a linear approximation to that 
density that has the correct initial and final values, as
well as the correct density and density derivative at $r_c$.
Thus the linear profile is a very good approximation to the
sun in the vicinity of the crossing point.  The MSW equations 
can be solved analytically for this wedge.  By extending the
wedge to $\pm \infty$ (dotted lines) and assuming adiabatic
propagation in these regions of unphysical density, one obtains
the simple Landau-Zener result discussed in the text.}
\end{figure}
  
That result can be simplified further
by observing that the nonadiabatic region is generally confined to 
a narrow region around $x_c$, away from the endpoints $x_i$ and $x_f$.  We  
can then extend the artificial profile to $x = \pm \infty$, as illustrated by  
the dashed lines in Fig. 5.  As the neutrino propagates adiabatically in the 
unphysical region $x < x_i$, the exact soluation in the physical region can be 
recovered by choosing the initial boundary conditions
\begin{eqnarray}
a_L (- \infty) &=& - a_\mu (- \infty) = \cos \theta_i e^{- i \int^{x_i}_{- 
\infty} \lambda (x) dx} \nonumber\\
a_H (- \infty) &=& a_e (- \infty) = \sin \theta_i e^{i \int^{x_i}_{- \infty} 
\lambda (x) dx} . 
\end{eqnarray}
That is, $|\nu (-\infty)\rangle$ will then adiabatically evolve to $|\nu 
(x_i)\rangle = |\nu_e\rangle$ as $x$ goes from $- \infty$ to $x_i$.  The 
unphysical region $x > x_f$ can be handled similarly.  

With some algebra a simple generalization of the adiabatic
result emerges that is valid for all $\delta m^2/E$ and $\theta_v$
\begin{equation}
P_{\nu_e} = {1 \over 2} + {1 \over 2} \cos 2 \theta_v \cos 2 \theta_i ( 1 - 
2P_{\rm {hop}}) 
\end{equation}
where P$_{\rm {hop}}$ is the Landau-Zener probability of hopping from the heavy mass
trajectory to the light trajectory on traversing the crossing
point.  For the linear approximation to the density~\cite{hlz,plz},
\begin{equation}
P^{\rm {lin}}_{\rm {hop}} = e^{- \pi \gamma_c/2} . 
\end{equation}
As it must by our construction, $P_{\nu_e}$ reduces to P$^{\rm 
{adiab}}_{\nu_e}$ for $\gamma_c \gg$ 1.   
When the crossing becomes nonadiabatic (e.g., $\gamma_c \ll 1$ ),
the hopping probability goes to 1, allowing the neutrino to
exit the sun on the light mass trajectory as a $\nu_e$, i.e., no conversion 
occurs. 

Thus there are two conditions for strong           
conversion of solar neutrinos:  there must be a level 
crossing (that is, the solar core density must be sufficient 
to render $\nu_e \sim \nu_H (x_i)$  when it is first
produced) and the crossing must be adiabatic.  The first
condition requires that $\delta m^2/E$ not be too large, and the 
second $\gamma_c \gsim$ 1.  The combination of these two constraints,
illustrated in Fig. 6, defines a triangle of interesting
parameters in the ${\delta m^2 \over E} - \sin^2 2\theta_v$ plane, as Mikheyev and Smirnov
first found.  A remarkable feature of this triangle
is that strong $\nu_e \to \nu_\mu$ conversion can occur for very small
mixing angles $(\sin^2 2 \theta \sim10^{-3}$), unlike the vacuum case. 

\begin{figure}[htb]
\psfig{bbllx=-1.5cm,bblly=0.0cm,bburx=15cm,bbury=22.0cm,figure=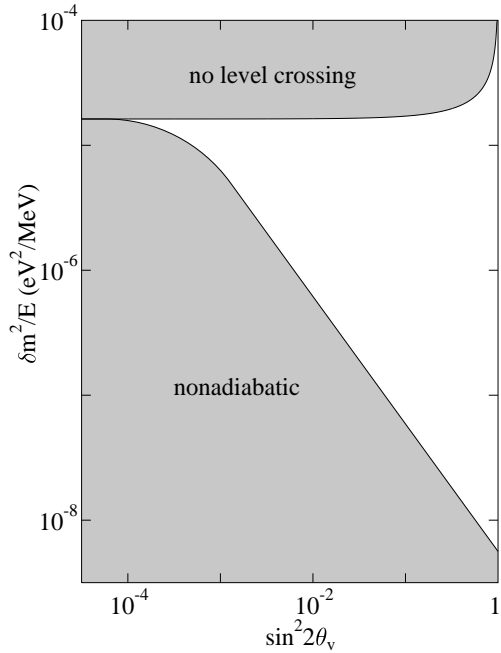,height=3.6in}
\caption{MSW conversion for a neutrino produced at the sun's
center.  The upper shaded region indices thoses $\delta m^2/E$
where the vacuum mass splitting is too great to be overcome
by the solar density.  Thus no level crossing occurs.  The
lower shaded region defines the region where the level crossing
is nonadiabatic ($\gamma_c$ less than unity).  The unshaded
region corresponds to adiabatic level crossings where strong
$\nu_e \rightarrow \nu_\mu$ will occur.}
\end{figure}
  
One can envision superimposing on Fig. 6 the spectrum of solar neutrinos, plotted as a 
function of ${\delta m^2 \over E}$ for some choice of $\delta m^2$.
Since Davis sees {\it some} solar neutrinos, the solutions must 
correspond to the boundaries of the triangle in Fig. 6.  The horizontal 
boundary indicates the maximum ${\delta m^2 \over E}$ for which the sun's 
central density is sufficient to cause a level crossing.  If a spectrum 
properly straddles this boundary, we obtain a result consistent with the 
Homestake experiment in which low energy neutrinos (large 1/E) lie above the 
level-crossing boundary (and thus remain $\nu_e$'s), but the high-energy 
neutrinos (small 1/E) fall within the unshaded region where strong conversion 
takes place.  Thus such a solution would mimic nonstandard solar models in 
that only the $^8$B neutrino flux would be strongly suppressed.  The diagonal 
boundary separates the adiabatic and nonadiabatic regions.  If the spectrum 
straddles this boundary, we obtain a second solution in which low energy 
neutrinos lie within the conversion region, but the high-energy neutrinos 
(small 1/E) lie below the conversion region and are characterized by $\gamma 
\ll 1$ at the crossing density.  (Of course, the boundary is not a sharp one, 
but is characterized by the Landau-Zener exponential).  Such a nonadiabatic 
solution is quite distinctive since the flux of  pp neutrinos, which is 
strongly constrained in the standard solar model and in any steady-state 
nonstandard model by the solar luminosity, would now be sharply reduced.  
Finally, one can imagine ``hybrid" solutions where the spectrum straddles both
the level-crossing (horizontal)
boundary and the adiabaticity (diagonal) boundary for small $\theta$,
thereby reducing the $^7$Be neutrino flux more than either the
pp or $^8$B fluxes. 

What are the results of a careful search for MSW solutions
satisfying the Homestake, Kamiokande/SuperKamiokande, and SAGE/GALLEX constraints?
One solution, corresponding to a region surrounding $\delta m^2 \sim 6 \cdot 10^{-6} $eV$^2$ 
and $\sin^2 2\theta_v \sim 6 \cdot 10^{-3}$, is the hybrid case described above.  It is commonly 
called the small-angle solution.  A second, large-angle solution
exists, corresponding to $\delta m^2 \sim 10^{-5} $eV$^2$ and $\sin^2 2 \theta_v 
\sim$ 0.6.  (Variations on these solutions include 
oscillations to sterile neutrinos, oscillations modified by 
regeneration as neutrinos pass through the earth (day/night effects),
and of course vacuum oscillations.)
These solutions can be distinguished by their characteristic
distortions of the solar neutrino spectrum.  The survival
probabilities $P_{\nu_e}^{\rm MSW}$(E) for the small- and large-angle parameters
given above are shown as a function of E in Fig. 7.

\begin{figure}[htb]
\psfig{bbllx=0.5cm,bblly=1.3cm,bburx=18cm,bbury=13.7cm,figure=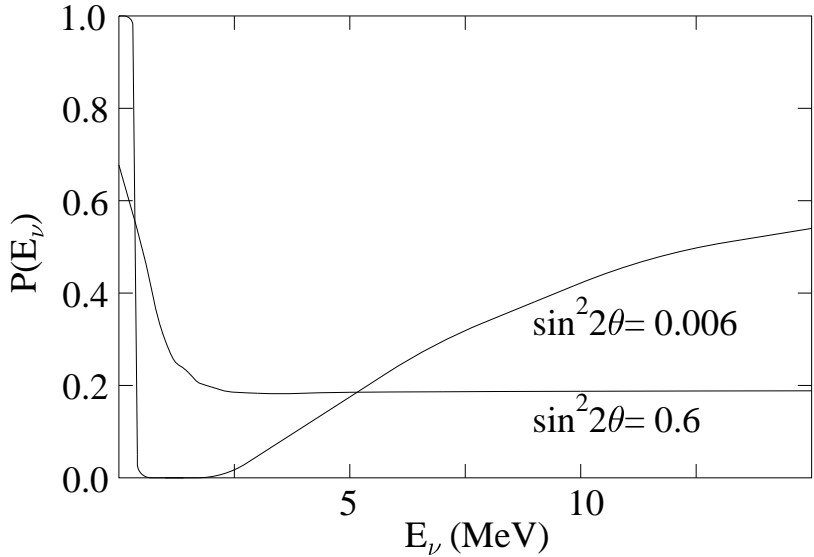,height=3.0in}
\caption{MSW survival probabilities P(E$_\nu$) for typical small
angle and large angle solutions.}
\end{figure}
  
The MSW mechanism provides a natural explanation for the
pattern of observed solar neutrino fluxes.  While it requires
profound new physics, both massive neutrinos and neutrino mixing
are expected in extended models.  The small-angle solution 
corresponds to $\delta m^2 \sim 10^{-5}$ eV$^2$, and thus is consistent with
$m_2 \sim$ few $\cdot 10^{-3}$ eV.  This is a typical $\nu_\tau$ mass in models
where $m_R \sim m_{\rm {GUT}}$.  This mass is also reasonably close to atmospheric neutrino values.  On the other hand, if it is the $\nu_\mu$  
participating in the oscillation, this gives $m_R \sim 10^{12}$ GeV
and predicts a heavy $\nu_\tau \sim$ 10 eV.  Such
a mass is of great interest cosmologically as it would have
consequences for supernova physics,
the dark matter problem, and the formation of large-scale structure. 

There are many interesting elaborations of the MSW effect not
discussed here, but treated in many papers:
spin-flavor oscillations induced by the solar magnetic field \cite{lim,ak}
(the mass difference between $\nu^L_e$ and a sterile $\nu^R_\mu$
is compensated by the matter effects); oscillations induced by
density fluctuations \cite{schaefer,krastev,zhang}; ``stochastic depolarization'' effects in
large random magnetic fields \cite{loreti}; etc.  
There are also interesting effects associated with three neutrinos:
if the solar neutrino problem is due to MSW $\nu_e-\nu_\mu$
oscillations, one might expect a $\nu_e-\nu_\tau$ crossing
at still higher densities.  This has led to many interesting
speculations about the role of the MSW mechanism in
supernova explosions and in supernova nucleosynthesis,
and about the possibility that $\nu_e - \nu_\tau$ oscillations governed
by small mixing angles might be best probed using the supernova
neutrino flux.
    
\section{Other Neutrino Mass Evidence and Implications}
The solar neutrino problem is not the most compelling evidence for
neutrino mass.  SuperKamiokande's analysis of the ratio of
muon-like to electron-like atmospheric neutrino events confirmed that a
dramatic anomaly exists.  The quantity studied is
\begin{equation}
R = {(N_\mu/N_e)_{DATA} \over (N_\mu/N_e)_{MC}}, 
\end{equation}
the measured ratio of muon-like to electron-like neutrino
events normalized to the expected ratio, based on Monte Carlo
calculations of the production and interaction of atmospheric neutrinos.
The SuperKamiokande results are \cite{skatm}
\begin{equation}
  R=0.63 \pm 0.03 ({\rm stat}) \pm 0.05 ({\rm syst}) 
\end{equation}
for sub-GeV events which were fully contained in the detector and
\begin{equation}
  R=0.65 \pm 0.05 ({\rm stat}) \pm 0.08 ({\rm syst}) 
\end{equation}
for fully- and partially-contained multi-GeV events. 
The results for $R$ are consistent among the four largest detectors
used for atmospheric neutrinos (SuperK, Soudan II, IMB, Kamiokande).
While this suggests neutrino oscillations, even stronger evidence
for new physics comes from measurements of $R$ as a function of the
zenith angle, $\Theta$, between the vertical and neutrino direction. A
down-going neutrino ($\Theta \sim 0^o$) travels through the atmosphere
above the detector (a distance of about 20 km), whereas an up-going
neutrino ($\Theta \sim 180^o$) has traveled through the entire Earth
(a distance of about 13000 km). Hence a measurement of number of
neutrinos as a function of the zenith angle yields information about
their numbers as a function of the distance traveled. 

\begin{figure}[t]
  \vspace{8pt} \centerline{\hbox{\epsfxsize=3 in 
\epsfbox[8 -3 502 485]{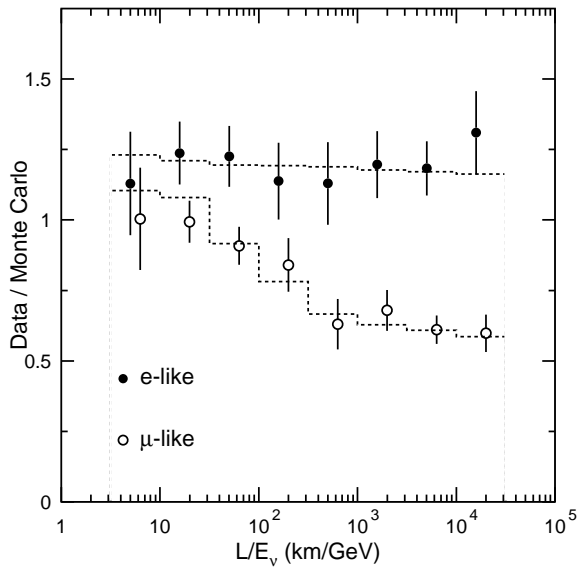}}}
\caption{The ratio of fully contained events measured
  at SuperKamiokande versus reconstructed $L/E_{\nu}$. The dashed
  lines show the expected shape for $\nu_{\mu} \leftrightarrow
  \nu_{\tau}$ oscillations with $\delta m^2 = 2.2 \cdot 10^{-3}$
  eV$^2$ and $\sin^2 2 \theta = 1$. } 
\vspace{8pt}
\end{figure}

The zenith angle dependence of the electron and muon fluxes \cite{skatm}
is shown in Fig. 8, plotted as a function of the reconstructed $L/E_{\nu}$.
The muon neutrino flux drops with increasing distance, while
the electron neutrino flux is approximately constant.
This behavior is consistent with $\nu_{\mu} \leftrightarrow
\nu_{\tau}$ oscillations.  Measurements of up-going muons
at Kamiokande \cite{katm} and
MACRO \cite{matm}, produced by high energy muon neutrino interactions
in the rock below the detectors, show a similar deficit.

It is often pointed out that such results provide convincing evidence because
the up/down difference in R is essentially self-normalizing:
only weak geomagnetic effects are expected to break the 
isotropy of cosmic ray interactions in the atmosphere.

Explanations of the SuperK anomaly in terms of $\nu_\mu \leftrightarrow
\nu_e$ oscillations conflict with reactor oscillation limits,
while the explanation $\nu_\mu \leftrightarrow \nu_{sterile}$ is
disfavored in fits.  The preferred solution is $\nu_\mu \leftrightarrow
\nu_\tau$ characterized by
\[ \delta m_{23}^2 \sim (4 \pm 2) \cdot 10^{-3} eV^2 \]
\[ 0.8 \lsim \sin^22\theta \lsim 1.0.\]
The best fit mixing angle is approximately maximal, an intriguing and surprising
result.

One can take the the square root of the atmospheric $\delta m^2$ to 
obtain $\sim 0.06$ eV, a minimum value for a neutrino mass.  This
immediately establishes a lower bound on the neutrino contribution
to dark matter of about 0.3\% the closure density, a value not 
too different from the mass evident in visible stars, a remarkable
result \cite{turner}.

There is one other indication of neutrino mass, 
the positive signal for $\bar{\nu}_\mu \rightarrow \bar{\nu}_e$
oscillations seen by the LSND group in a beam-stop experiment
at Los Alamos \cite{lsnd}.  The experiment uses a 52,000 gallon tank of 
mineral oil and liquid scintillator, instrumented with 1220
phototubes.  A neutrino event $\bar{\nu}_e + p \rightarrow 
n + e^+$ is detected by the coincidence of the positron and 
subsequent 2.2 MeV $\gamma$ ray from neutron capture,
$n + p \rightarrow d + \gamma$.  The signal is consistent 
with $\bar{\nu}_\mu \rightarrow \bar{\nu}_e$ oscillations 
in a narrow band that includes the ranges $\delta m^2 \sim$
0.2 - 2.0 eV$^2$ and $\sin^22\theta \sim 0.03 - 0.003$.
A similar experiment at the Rutherford Laboratory, KARMEN, 
sees no oscillations, but has lower sensitivity \cite{karmen}.  A recent 
combined analysis \cite{eitel} of the two experiments 
lowers the confidence level of the oscillation claim, but
finds a parameter region consistent with both experiments.
An improved experiment at Fermilab has been approved and
should yield results in 2002.

These results have some interesting implications.  For example, we discussed
the quadratic seesaw relation earlier, $m_{\mathrm{light}}\sim {m_D^2
\over m_R}$.  If we use the atmospheric $\delta m^2$ 
as a rough guide to the $\nu_\tau$ mass (or more correctly the
$\nu_3$ mass, given the large atmospheric neutrino mixing angle),
$\nu_3 \sim 0.1$ eV, and adopt for $m_D$ the corresponding 
third-generation quark mass, $m_D \sim m_{top} \sim$ 180 GeV,
one obtains $m_R \sim 0.3 \cdot 10^{15}$ GeV.  This is a value
reasonable close to the supersymmetric grand unified scale 
of $\sim 10^{16}$ GeV, a startling result.  It has inspired some
to hope that current neutrino results are giving us our first
glimpse of physics at the GUT scale.  

One puzzling aspect of atmospheric, solar, and LSND neutrino
results is that they require three independent $\delta m^2$s.
That is, they do not respect the relation
\begin{equation}
\delta m^2_{21} + \delta m^2_{13} + \delta m^2_{32} = 0.
\end{equation}
Thus either one of more of the experiments must be attributed
to some phenomenon other than neutrino oscillations, or a
fourth neutrino is required.  That neutrino must be sterile to
avoid constraints imposed by the known width of the $Z_0$.

\section{Outlook}

The argument that the solar neutrino problem is due to neutrino
oscillations is, in a sense, circumstantial: this conclusion is
derived from combining several experiments, no one of which 
requires new particle physics.
There is no direct observation of new physics analogous to the
zenith angle dependence of the SuperKamiokande atmospheric
results.  For this reason there is great interest in a new
experiment now taking data in the Creighton nickel mine in
Sudbury, Ontario, 6800 feet below the surface.  The Sudbury
Neutrino Observatory (SNO) has a central acrylic vessel filled
with one kiloton of very pure (99.92\%) heavy water, surrounded
by a shield of 7.5 kilotons of ordinary water.  SNO can detect
electron neutrinos through the charged current reaction
\begin{equation}
\nu_e + d \rightarrow p + p + e^-
\end{equation}
The Cerenkov light from the outgoing electron is then recorded
in the array of 9800 phototubes that surround SNO's central vessel.
The spectrum of produced electrons is quite hard, making 
reconstruction of the energy of the $\nu_e$ easier than in the
case of neutrino-electron elastic scattering.  Thus the
experimenters may be able to detect distortions of the
neutrino spectrum resulting from the MSW effect.

SNO will also study the neutral current reaction
\begin{equation}
\nu_x(\bar{\nu}_x) + d \rightarrow \nu_x(\bar{\nu}_x) + p + n
\end{equation}
by detecting the produced neutron either through $(n,\gamma)$
reactions on salt dissolved in the heavy water or in 
$^3$He proportional counters.  In this way the experimenters
will obtain an integral measurement of the flux of active
neutrinos, independent of flavor.  Thus a neutral current
signal clearly larger than the corresponding $\nu_e$ signal
would show that heavy-flavor neutrinos comprise a portion of
the solar neutrino flux, providing definite proof of new 
physics.  

The SNO collaboration is expected to make its first announcement
of results for the charged current reaction as early as summer,
2000.  In future years, as the SNO neutral and charged current
results become precise and as SuperKamiokande continues to amass
data, the nature of the solar neutrino problem should become
much clearer.  The hope is that these results, in combination 
with other solar neutrino results (Borexino, GNO, iodine),
with new atmospheric and (possibly) supernova neutrino 
measurements, and with precision tests of oscillations at accelerators
and reactors, will allow us to completely characterize the
neutrino mass matrix, providing a window on physics well beyond
the standard model.

This work was supported in part by the US Department of Energy.

\end{document}